\documentclass[amsmath, preprint, floatfix]{revtex4-2}
\usepackage{physics}
\usepackage{siunitx}
\usepackage{graphicx}
\usepackage{microtype}

\begin{document}

\title{Timing and energy stability of resonant dispersive wave emission in gas-filled hollow-core waveguides}

\author{Christian Brahms}
\email[Corresponding author: ]{c.brahms@hw.ac.uk}
\author{John C. Travers}
\affiliation{School of Engineering and Physical Sciences, Heriot-Watt University, Edinburgh, EH14 4AS, UK}

\begin{abstract}
We numerically investigate the energy and arrival-time noise of ultrashort laser pulses produced via resonant dispersive wave emission in gas-filled hollow-core waveguides under the influence of pump-laser instability. We find that for low pump energy, fluctuations in the pump energy are strongly amplified. However, when the generation process is saturated, the energy of the resonant dispersive wave can be significantly less noisy than that of the pump pulse. This holds for a variety of generation conditions and while still producing few-femtosecond pulses. We further find that the arrival-time jitter of the generated pulse remains well below one femtosecond even for a conservative estimate of the pump pulse energy noise, and that photoionisation and plasma dynamics can lead to exceptional stability for some generation conditions. By applying our analysis to a scaled-down system, we demonstrate that our results hold for frequency conversion schemes based on both small-core microstructured fibre and large-core hollow capillary fibre.
\end{abstract}

\maketitle

\section{Introduction}
\begin{figure}
    \includegraphics[width=6in]{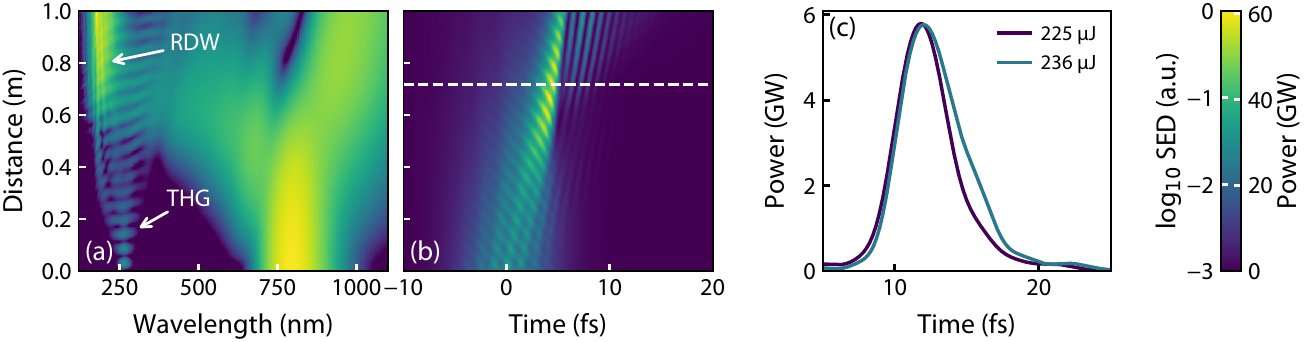}
    \caption{Example simulation of soliton self-compression and RDW emission. (a) Evolution of the pulse spectrum along the waveguide when pumping a helium-filled hollow-capillary fibre with \SI{125}{\micro\meter} core radius and \SI{1}{\m} length with \SI{7.5}{\fs} pulses containing \SI{225}{\micro\joule} of energy. The labels point out the spectral bands of RDW emission and third-harmonic generation (THG). (b) Corresponding evolution of the pulse envelope in time. The white dashed line indicates the calculated point of maximal self-compression (see section \ref{subsec:timing}). (c) Pulse profile of the RDW at the end of the waveguide, filtered with \SI{15}{\percent} relative bandwidth around \SI{180}{\nm}, for two different pump energies.}
    \label{fig:example}
\end{figure}

Resonant dispersive wave (RDW) emission in gas-filled hollow-core waveguides is a promising source of tuneable few-femtosecond pulses from the vacuum ultraviolet (VUV) to the near infrared \cite{joly_bright_2011, mak_tunable_2013,ermolov_supercontinuum_2015, travers_high-energy_2019, brahms_infrared_2020}. It is based on resonant energy transfer to a phase-matched spectral band during soliton self-compression of a laser pulse \cite{erkintalo_cascaded_2012,akhmediev_cherenkov_1995}. This is fundamentally a dynamical and strongly nonlinear process: anomalous group-velocity dispersion continually compensates the positive chirp induced by the optical Kerr effect, compressing the pump pulse as it propagates and increasing the intensity, which leads to more rapid spectral broadening and, ultimately, the formation of sub-cycle pulses (see Fig.~\ref{fig:example}) \cite{travers_high-energy_2019, brahms_infrared_2020}. This feedback loop lies at the heart of the unique capabilites offered by RDW emission---most notably, wavelength tuneability and the generation of pulses significantly shorter than the pump pulse. However, it may also make the process more sensitive to changes in the pump pulse than, for instance, simple spectral broadening in the normal-dispersion regime \cite{heidt_pulse_2010}.

The noise properties of soliton effects have been studied in detail in the context of supercontinuum generation in solid-core fibres, using both numerical and experimental methods and considering practical as well as fundamental limitations \cite{nakazawa_coherence_1998, dudley_coherence_2002,corwin_fundamental_2003, nicholson_cross-coherence_2004,genty_complete_2011}. Experimental studies of soliton self-compression in gas-filled hollow-core waveguides have found that both the pulse energy and absolute phase of self-compressed few-cycle pulses can be stabilised to the same degree as in normal-dispersion systems \cite{kottig_efficient_2020,ermolov_carrier-envelope-phase-stable_2019}. Ultraviolet RDW emission has also been investigated purely in terms of the spectral intensity noise, which was found to be very low for correctly chosen generation parameters \cite{travers_high-energy_2019} but depend strongly on the pulse-energy noise of the driving laser \cite{adamu_noise_2020,smith_low-noise_2020}. However, these works investigate only one or at most a few different combinations of generation parameters (gas pressure, nominal pump energy and duration), so that the effect of changing, for instance, the RDW wavelength remains unknown.

Since it enables the efficient generation of few-femtosecond pulses across the vacuum and deep ultraviolet, where many chemical compounds have absorption resonances, RDW emission holds particular promise in the field of ultrafast spectroscopy. Experiments in this field require a high degree of timing stability between two or more pulses used to excite, control and probe ultrafast dynamics in a sample. In this respect, the dynamical nature of the RDW emission process poses a challenge. In contrast to conventional frequency-conversion schemes (for instance harmonic generation), the dispersive wave is not automatically locked to the pump pulse. Instead, it is generated around the point of maximal self-compression---the location of which depends sensitively on the pump pulse---and then propagates under the influence of both linear dispersion and further nonlinear interactions with the soliton. The variable propagation length in combination with a difference in group velocity means that the arrival time of the RDW pulse at the waveguide exit is coupled to the pump pulse energy and duration [see Fig.~\ref{fig:example}(c)].

Here we use numerical simulations to study the noise properties of RDW emission for a range of generation conditions, considering fluctuations in the generated energy, the arrival time, and the central wavelength. We find that for low pump energy, pump pulse noise is strongly amplified. However, at higher energy, where the generation is saturated, the energy noise of the RDW pulse can be significantly lower than that of the pump pulse. This effect is consistent across different gas pressures (corresponding to different RDW wavelengths) and for both a constant gas pressure and a longitudinal pressure gradient, which can be used for direct delivery of compressed RDW pulses to a vacuum system \cite{brahms_direct_2019, brahms_resonant_2020}. We further find that, even for a conservative estimate of the pump pulse noise, the arrival-time jitter of the RDW pulse remains well below 1 femtosecond. Higher gas pressures (longer RDW wavelengths) generally perform better in terms of timing jitter, but photoionisation dynamics lead to exceptional stability for very low pressure. Comparison between energy-scaled systems reveals that the energy and timing noise characteristics we identify here apply to RDW emission in both small-core microstructured fibres and large-core hollow-capillary fibres.

\section{Numerical model}
\subsection{Propagation equation}
To accurately simulate the nonlinear propagation of laser pulses through a gas-filled hollow-core waveguide, we use a carrier-resolved multi-mode model which includes the effect of gas and waveguide dispersion, confinement loss, the optical Kerr effect, and photoionisation and plasma dynamics. This model achieves quantitative agreement with experimental results without any fitting parameters \cite{travers_high-energy_2019}. We numerically solve the multi-mode unidirectional pulse propagation equation \cite{kolesik_nonlinear_2004,travers_high-energy_2019,tani_multimode_2014},
\begin{equation}
    \label{eq:modeprop}
    \partial_z E_m(\omega,z)=\left(i\beta_m(\omega, z)-i\frac{\omega}{v(z)}-\frac{1}{2}\alpha_m(\omega, z)\right)E_m(\omega,z) + i\frac{\omega}{4}P_m^\mathrm{nl}(\omega,z) \,,
\end{equation}
where $E_m(\omega,z)$ is the modal electric field in mode $m$ as a function of angular frequency $\omega$ and propagation distance $z$, $\alpha_m(\omega, z)$ and $\beta_m(\omega, z)$ are the frequency-dependent attenuation and propagation constants of mode $m$, respectively, $v(z)$ is the velocity of the reference frame (chosen as the group velocity at the central wavelength of the pump pulse in the fundamental mode), and $P_m^\mathrm{nl}(\omega,z)$ is the nonlinear polarisation. We consider propagation in the radially symmetric and linearly polarised hybrid modes of the capillary, HE$_{1m}$, for which the attenuation and propagation constants are obtained from the capillary model \cite{marcatili_hollow_1964} as
\begin{equation}
    \beta_m(\omega, z) = \frac{\omega}{c}\sqrt{n_\mathrm{gas}^2(\omega, z) - \frac{c^2u_{1m}^2}{a^2\omega^2}}\,,\qquad \alpha_m(\omega, z) = \frac{c^2 u_{1m}^2}{a^3\omega^2} \frac{\nu(\omega, z)^2+1}{\sqrt{\nu(\omega, z)^2 -1}}\,,
\end{equation}
where $c$ is the speed of light in vacuum, $u_{1m}$ is the $m^\mathrm{th}$ zero of the Bessel function of the first kind $J_0$, $a$ is the core radius of the capillary, $n_\mathrm{gas}(\omega, z)$ is the refractive index of the filling gas, and $\nu(\omega, z)$ is the ratio of the refractive index of the cladding and $n_\mathrm{gas}$, $\nu(\omega, z) = n_\mathrm{clad}(\omega, z)/n_\mathrm{gas}(\omega, z)$. $n_\mathrm{gas}(\omega, z)$ is calculated by scaling the refractive index, obtained from Sellmeier expansions \cite{ermolov_supercontinuum_2015,borzsonyi_dispersion_2008}, to the correct pressure using the pressure-dependent number density $\rho(z)$ \cite{bell_pure_2014}. For the case of a decreasing pressure gradient for direct coupling to vacuum, the pressure along the waveguide $p(z)$ is described by
\begin{equation}
    p(z) = p_0\sqrt{1-\frac{z}{L}}\,,
\end{equation}
where $p_0$ is the fill pressure and $L$ is the length of the waveguide.

The modal nonlinear polarisation $P_m^\mathrm{nl}(\omega, z)$ is obtained from the real-space polarisation $P^\mathrm{nl}\!\left(t, r, \theta, z\right)$ by projecting onto the normalised modal fields $\hat{e}_m$ and taking the Fourier transform,
\begin{equation}
    \label{eq:modeproj}
    P_m^\mathrm{nl}(\omega,z) = \int_{-\infty}^{+\infty}\!\!\mathrm{d}\omega \int_0^{2\pi\!\!}\mathrm{d}\theta\int_0^a \!\!r\mathrm{d}r\,\hat{e}_m^*(r,\theta)\cdot P^\mathrm{nl}(t,r,\theta,z) \mathrm{e}^{i\omega t}\,.
\end{equation}
To calculate the real-space polarisation, in turn, the real-space field $E\!\left(t,r,\theta,z\right)$ is constructed from the modal fields $E_m(\omega, z)$,
\begin{equation}
    \label{eq:mode_expansion}
    E\!\left(t,r,\theta,z\right) = \frac{1}{2\pi}\int_{-\infty}^{+\infty}\!\!\mathrm{d}\omega \,\mathrm{e}^{-i\omega t} \sum_m \hat{e}_m(r,\theta)E_m(\omega, z)\,.
\end{equation}
The real-space polarisation consists of two parts capturing the effects of third-order (Kerr) nonlinearity and photoionisation and plasma, respectively:
\begin{equation}
    P^\mathrm{nl}(t,r,\theta,z) = P^\mathrm{Kerr}\!\left(t,r,\theta,z\right) + P^\mathrm{ion}\!\left(t,r,\theta,z\right)\,.
\end{equation}
The Kerr term is given by
\begin{equation}
    P^\mathrm{Kerr}\!\left(t,r,\theta,z\right)=\rho(z)\epsilon_0\gamma^{(3)}E^3\!\left(t,r,\theta,z\right)\,,
\end{equation}
where $\epsilon_0$ is the permittivity of free space and $\gamma^{(3)}$ is the third-order hyperpolarisability of the gas, calculated from reference values for the third-order susceptibility \cite{lehmeier_nonresonant_1985}. To suppress third-harmonic generation, the Kerr term can be altered to
\begin{equation}
    P^\mathrm{Kerr}\!\left(t,r,\theta,z\right)=\frac{3}{4}\rho(z)\epsilon_0\gamma^{(3)}\left\vert\hat{E}\!\left(t,r,\theta,z\right)\right\vert^2 E\!\left(t,r,\theta,z\right)\,,
\end{equation}
where $\hat{E}$ is the analytic representation of the field, obtained from $\hat{E} = E + i\mathcal{H}\left[E\right]$ with $\mathcal{H}$ denoting the Hilbert transform. The photoionisation term is given by \cite{geissler_light_1999}
\begin{equation}
    P^\mathrm{ion}\!\left(t,r,\theta,z\right) = I_p\int_{-\infty}^t \!\!\mathrm{d}t'\frac{\partial_{t'} \rho_\mathrm{e}(t', r,\theta,z)}{E\!\left(t',r,\theta,z\right)} + \frac{e^2}{m_\mathrm{e}}\int_{-\infty}^{t}\!\!\mathrm{d}t'\int_{-\infty}^{t'} \!\!\mathrm{d}t'' \rho_\mathrm{e}(t'',r,\theta,z)E\left(t'',r,\theta,z\right)\,,
\end{equation}
where $I_p$ is the ionisation potential of the gas, $\rho_\mathrm{e}(t, r, \theta, z)$ is the density of free electrons, $e$ is the elementary charge, and $m_\mathrm{e}$ is the mass of the electron. $\rho_\mathrm{e}(t, r, \theta, z)$ is calculated from
\begin{equation}
    \rho_\mathrm{e}(t, r,\theta,z) = \rho(z)\left(1 - \exp \left\{-\int_{-\infty}^t\!\!\mathrm{d}t'w\!\left(\left\vert E\!\left(t',r,\theta,z\right)\right\vert\right)\right\}\right)\,,
\end{equation}
where $w(E)$ is the ionisation rate after Perelomov \emph{et al.}~\cite{perelomov_ionization_1966}.

We solve eq.~\ref{eq:modeprop} with the Dormand-Prince method (an adaptive 4(5)$^\mathrm{th}$-order Runge-Kutta method) \cite{dormand_family_1980} after transforming the propagation equation into the interaction picture \cite{hult_fourth-order_2007}. The modal projection in eq.~\ref{eq:modeproj} is carried out using a p-adaptive cubature method to integrate over the core of the capillary. We consider the first four hybrid modes of the capillary (HE$_{11}$ to HE$_{14}$). The grid used in the simulations extends from \SI{100}{\nm} to \SI{3}{\micro\meter} in wavelength with a time window of \SI{450}{\femto\second}, resulting in 4096 samples. To accurately calculate the nonlinear polarisation including third-harmonic generation and the plasma response, the field is resampled onto a finer grid with a time resolution of \SI{55}{\atto\second}. Convergence tests show that this is sufficient resolution for all parameters we consider here. 

\subsection{Statistical sampling}
The most direct way of modelling the noise properties of RDW emission is to simply repeatedly simulate the process while randomly choosing input parameters according to a given model of the pump pulse noise. This approach is commonly used to analyse the coherence properties of supercontinuum generation \cite{dudley_coherence_2002} and has previously been applied to RDW emission in gas-filled hollow-core waveguides to investigate the effects of quantum noise \cite{mak_tunable_2013} and pump pulse energy fluctuations \cite{adamu_noise_2020} for a fixed combination of waveguide and nominal pump pulse parameters. However, for a more comprehensive study of the influence of various experimental parameters on the RDW noise using an accurate multi-mode model, the very large number of individual propagation simulations required makes the computational cost of this method prohibitive.

\begin{figure}
    \centering
    \includegraphics[width=6in]{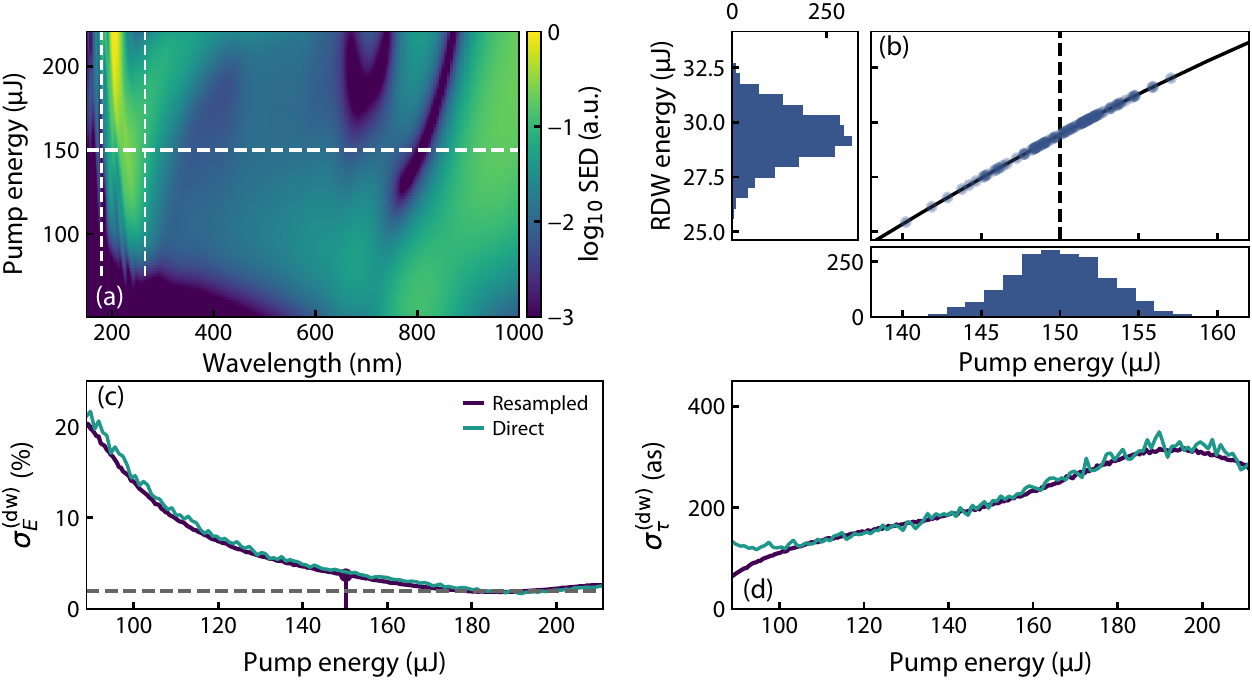}
    \caption{(a) Simulated output spectra for a \SI{1}{\meter} long HCF with \SI{125}{\micro\meter} core radius at a helium pressure of \SI{2.1}{\bar} pumped with \SI{7.5}{\fs} pulses centred at \SI{800}{\nm}. (b) Illustrative example of the resampling approach to extracting noise characteristics: an interpolant of the RDW energy as a function of pump energy is resampled at random points with a standard deviation of \SI{2}{\percent} around \SI{150}{\micro\joule} (bottom histogram), creating a corresponding spread in the RDW energy (left histogram). (c, d) Comparison between directly sampled and resampled methods to model RDW emission noise for (c) total energy and (d) arrival time of the RDW pulse when and using a pump energy standard deviation of \SI{2}{\percent}. Direct sampling uses 500 independent propagation simulations for each mean pump energy whereas resampling uses 10000 samples from interpolated data.}
    \label{fig:direct_vs_fine}
\end{figure}

To reduce the computational cost, here we take a different approach based on re-sampling detailed parameter scans. Instead of running full simulations for randomly chosen input parameters, we run a fixed set of simulations with small increments in those same parameters [see Fig.~\ref{fig:direct_vs_fine}(a)]. We then extract the output quantities of interest---for instance, the energy of the RDW pulse---and create an interpolant for each quantity. This allows us to analyse the output noise by randomly choosing parameter values at which to sample the interpolant [see Fig.~\ref{fig:direct_vs_fine}(b)], which is several orders of magnitude faster than running a full simulation.

Assuming that environmental conditions (temperature, laboratory air pressure and humidity) and waveguide parameters (length, transverse dimensions, gas fill pressure and species) only vary on timescales much longer than an experiment, the two most important sources of noise are instabilities in the energy and duration of the pump pulse. However, two additional instabilities have to be taken into account. Firstly, in most pump laser systems the absolute phase, or carrier-envelope phase (CEP), of the pump pulse varies completely randomly from shot to shot. Secondly, quantum shot noise is unavoidable; since soliton self-compression is intimately related to modulational instability, which is seeded from such noise \cite{agrawal_nonlinear_2013}, it is important to assess its influence.

We validate our resampling approach by comparing it to the direct method for a representative set of parameters. Figure \ref{fig:direct_vs_fine}(c) and (d) show the results of this comparison for the total energy and arrival time of the dispersive wave generated in a \SI{1}{\meter} long HCF with \SI{125}{\micro\meter} core radius at \SI{2.1}{\bar} of helium pressure when pumping with transform-limited \SI{7.5}{\fs} Gaussian pulses centred at \SI{800}{\nm}. For each mean energy, the direct approach uses 500 independent simulations with an input energy that is randomly chosen from a Gaussian distribution with \SI{2}{\percent} standard deviation around the respective mean. These simulations also include uniformly random CEP and the effect of quantum shot noise using the one-photon-per-mode model \cite{dudley_coherence_2002}. We choose a fixed spectral window to capture the whole spectral band of the RDW; for the results in Fig.~\ref{fig:direct_vs_fine} the window spans \SIrange{185}{265}{\nm}. The RDW energy is extracted by integrating the filtered spectral energy density and summing over the waveguide modes, and the arrival time is extracted by transforming the filtered frequency-domain field to the time domain and calculating the first moment (centre of mass) of the resulting pulse. In our resampling approach, we run simulations with 800 uniformly spaced pump energies ranging from \SIrange{80}{220}{\micro\joule} (extending past the range of interest shown in Fig.~\ref{fig:direct_vs_fine}(c) and (d) to avoid edge artefacts) with fixed CEP and no shot noise. We then apply the same analysis to extract the RDW energy and arrival time and resample interpolants of the resulting data for a range of mean pump energy values from \SIrange{110}{210}{\micro\joule} using the same standard deviation of \SI{2}{\percent}.

The excellent agreement between the two methods demonstrates that the resampling approach can be used to analyse the noise properties of RDW emission. The small discrepancy in the timing jitter at pump energies below \SI{100}{\micro\joule} is most likely due to CEP-dependent interference between the RDW and the third harmonic (see section \ref{ssec:energy_noise}). The simulations for the resampling approach took 29 CPU hours as compared to over 1600 hours for the direct method, making the study of wide parameter ranges feasible.

Importantly, the agreement also shows that CEP fluctuations and quantum shot noise do not significantly influence the results in the parameter range we consider here. If this were not the case, the resampling approach would be invalid, since quantum noise in particular cannot be modelled in this way. That quantum noise has no effect is to be expected from previous numerical studies of RDW emission, which found that the process is fully coherent (insensitive to quantum noise) for typical parameters \cite{mak_tunable_2013}. In this context, it is important to note the difference between the coherence of any given real-world source and the generation process itself. In particular, a frequency-conversion or supercontinuum \emph{source} can be incoherent---that is, exhibit random shot-to-shot phase fluctuations---due to noise in the pump source, even if the \emph{underlying process} is fully coherent. In fact, because the majority of laser systems exhibit random CEP, most practical sources are incoherent in the strictest sense. The key distinction is that while pump-source noise can be mitigated by technical means, incoherence due to quantum noise is a fundamental limitation inherent to the process.

The use of a fixed frequency window makes the above comparison straightforward, but to capture the whole band of the RDW at all pump energies, it has to be more broadband than can realistically be achieved with filtering optics. For the following analysis, we instead use a variable window function with a fixed relative bandwidth of \SI{15}{\percent}. For each mean energy (each point on the horizontal axis in Fig.~\ref{fig:direct_vs_fine}) we find the spectral location of the RDW by calculating the first moment of the square of the spectral energy density in the ultraviolet. We then create a separate interpolant for each mean energy with windows centred on these wavelengths. This approach emulates the limitations of existing filtering optics while also taking into account the energy dependence of the RDW wavelength \cite{joly_bright_2011}, and thus avoids artefacts due to the RDW shifting into and out of the filter pass-band \cite{adamu_noise_2020,smith_low-noise_2020}. Since the generation parameters are usually chosen such that the RDW spectrum fits within the pass-band of the optics, it also mirrors experimental reality.

\section{Results and Discussion}
\subsection{Energy noise amplification}
\label{ssec:energy_noise}
\begin{figure}
    \centering
    \includegraphics[width=6in]{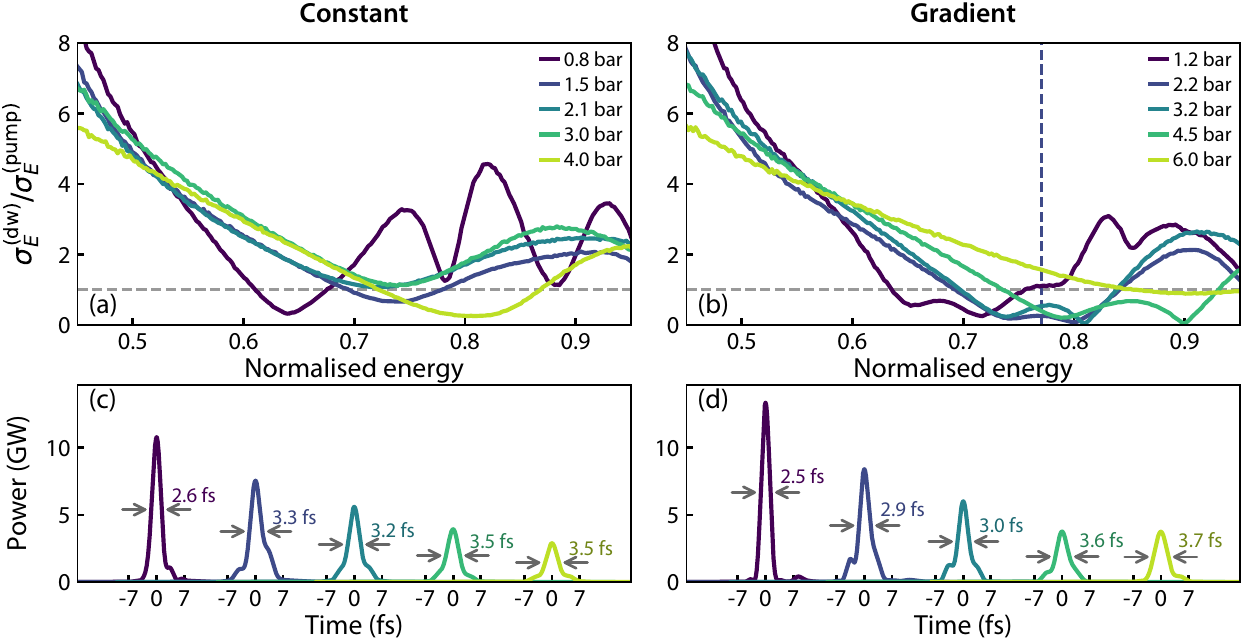}
    \caption{(a, b) Energy noise amplification of RDW emission as a function of mean pump energy in a \SI{1}{\m} long HCF with \SI{125}{\micro\meter} core radius filled with different pressures of helium at constant pressure (a) or with a decreasing gradient (b) and driven with \SI{7.5}{\fs} pulses at \SI{800}{\nm}. Horizontal dashed lines indicate a noise amplification of $1$, that is, the same noise in the pump and the dispersive wave. The vertical dashed line in (b) indicates the energy for which direct simulations are shown in Fig.~\ref{fig:min_direct}. (c, d) RDW pulses at the mean energy at which the lowest energy noise is observed, filtered with \SI{15}{\percent} relative bandwidth.}
    \label{fig:RDWenergy_std}
\end{figure}
Figure \ref{fig:RDWenergy_std}(a) and (b) show the energy noise amplification of RDW emission---the standard deviation of the RDW energy, $\sigma^\mathrm{(dw)}_E$, normalised by the that of the pump, $\sigma^\mathrm{(pump)}_E$. The pump energy standard deviation is chosen as \SI{2}{\percent}, however the noise amplification is insensitive to this choice even for values up to \SI{10}{\percent}. The gas pressures are chosen so as to generate a dispersive wave at a wide range of wavelengths from $\sim\!\SI{140}{\nm}$ to $\sim\!\SI{320}{\nm}$. For the decreasing gradient, the pressures are increased by a factor of $3/2$ \cite{brahms_resonant_2020}. The maximum energy is chosen so as to avoid excessive photoionisation effects at the lowest pressure we consider here. To compare results with different gas pressures, the pump energy is scaled by the pressure and normalised to this maximum energy. In this way, a particular value of the normalised energy corresponds to a fixed pressure-energy product. The energy range shown in Fig.~\ref{fig:RDWenergy_std} spans from the approximate normalised energy at which RDW emission begins to appear (0.45) to just below the maximum (1.0) to avoid edge effects. For \SI{0.8}{\bar} static pressure this range is \SIrange{260}{555}{\micro\joule}, whereas for \SI{4}{\bar} it is \SIrange{52}{111}{\micro\joule}. Note that these ranges cannot be arbitrarily scaled to any other system; they depend on several other parameters, including the waveguide dimensions, the gas species, and the pump pulse duration and wavelength.

The first important result is that the energy of the RDW generally becomes more stable as the mean energy is increased. The second is that, surprisingly, the energy noise of the RDW can be \emph{lower} than that of the pump pulse. In the extreme case in our data, the standard deviation in the RDW energy is a factor of 25 lower than that in the pump pulse energy. This goes counter to the conventional assumption that nonlinear optical processes amplify noise. The underlying physical reasons for this dramatic improvement in noise performance are not completely clear from our simulations. However, it is likely that saturation effects play an important role: at low pump energy, the pump pulse maximally self-compresses at or just before the end of the waveguide, and the rapid transfer of energy to the RDW band---which occurs around the self-compression point---is interrupted by the pulse exiting the waveguide. A small change in pump energy leads to earlier or later self-compression and thus has a large effect on how much energy is converted before the waveguide end. This amplifies the noise in the pump pulse. At high pump energy, the pump pulse maximally self-compresses well before the end of the waveguide, and conversion to the RDW band has ceased by the time the pulse reaches the end, making the process less sensitive. In addition, self-compression with more energy leads to more structure in the pulse at the point of RDW emission, which reduces the conversion efficiency \cite{chen_nonlinear_2002,travers_ultrafast_2011}. This may act as a second saturation mechanism leading to lower noise. Figure \ref{fig:RDWenergy_std}(c) and (d) show the temporal profiles of the filtered RDW pulse at the point of lowest energy noise in Fig.~\ref{fig:RDWenergy_std}(a) and (b), respectively. Their short duration shows that, to obtain good noise performance, RDW emission does not have to be driven so deep into saturation that dispersive effects distort the generated pulse.

The effect is reversed when the energy is increased beyond a certain point and the energy noise rises again. This is likely due to a combination of factors, including the increased effect of (highly nonlinear) photoionisation and plasma dynamics as well as coupling into higher-order modes. This is particularly clear in the data at the lowest pressure in the constant-pressure case, where the noise rises dramatically after the minimum.

The results with constant-pressure fill and a decreasing gradient are broadly similar, but the use of a gradient presents several advantages: firstly, the best energy noise achieved is lower than in the constant-pressure case for most gas pressures. Secondly, the energy noise is low for a larger range of pump energies, making such a system more flexible. Finally, the optimal energy noise is generally obtained at higher energy, which means that a gradient-filled system can generate more RDW energy before nonlinear instabilities degrade the noise performance. This is reflected in the higher peak power of the pulses in Fig.~\ref{fig:RDWenergy_std}(d).

Because the wavelength of RDW emission depends on the pump pulse energy \cite{joly_bright_2011}, the noise characteristics depend on the choice of spectral filter function. In the extreme case, placing the window on the edge of the RDW band results in very large relative energy noise as the spectral wings of the pulse shift into and out of the filter pass-band, even if the energy and spectrum of the RDW are stable \cite{adamu_noise_2020}. Experimentally, this is equivalent to detuning the RDW wavelength with respect to the filtering optics used. In our simulations we always find an energy at which the RDW energy noise is comparable to or better than the pump noise, irrespective of the chosen filtering bandwidth, as long as the window is centred on the spectral band of the RDW.

\begin{figure}
    \centering
    \includegraphics[width=6in]{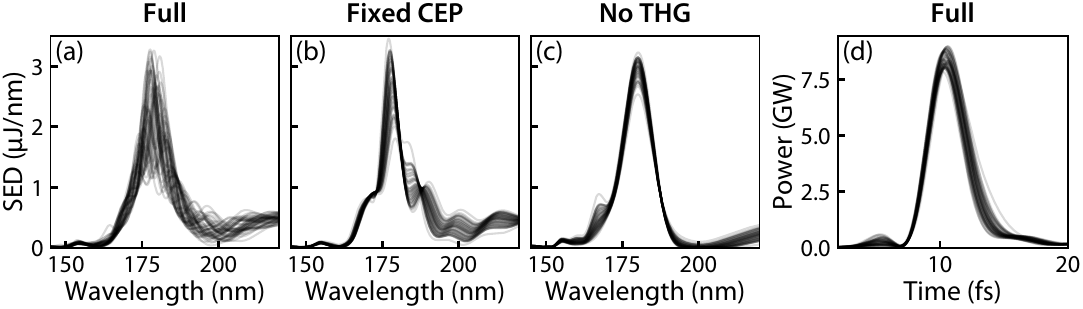}
    \caption{(a) Overlay of 50 individually simulated RDW spectra in a decreasing gradient from \SI{2.2}{\bar} to vacuum with a nominal pump pulse energy of \SI{240}{\micro\joule}, \SI{2}{\percent} energy noise, and including uniformly random CEP as well as quantum noise. (b) Same as (a) but with fixed CEP. (c) Same as (a) but with third-harmonic generation (THG) removed from the Kerr term in the nonlinear polarisation. (d) Temporal profile of the RDW pulse for the spectra shown in (a).}
    \label{fig:min_direct}
\end{figure}
To verify that the unexpectedly low noise of the RDW energy is not an artefact of our analysis, in Fig.~\ref{fig:min_direct} we show simulations using the direct approach discussed above. Here we overlay 50 shots with low-opacity lines to build a direct picture of the stability. The parameters are a gradient with \SI{2.2}{\bar} fill pressure and a nominal pump energy of \SI{240}{\micro\joule}, which corresponds to the centre of the low-noise region around a normalised energy of 0.75 [see vertical dashed line in Fig.~\ref{fig:RDWenergy_std}(b)]. For the complete model [Fig.~\ref{fig:min_direct}(a)], the spectrum appears very unstable, which would seem to contradict our earlier findings. However, the total energy noise is significantly lower than the fluctuations in spectral energy density. Closer inspection shows that the spectrum contains interference between the RDW and the third harmonic of the self-compressed sub-cycle pulse. This acts as an $f$-$3f$ interferometer encoding the CEP in the phase of the fringes \cite{genty_harmonic_2008-1}. Since the CEP is random in these simulations, this explains the strong instability. The same set of simulations with fixed CEP [Fig.~\ref{fig:min_direct}(b)] yields much lower noise in the spectral energy density, though it is still higher than would be expected from the extremely low overall energy noise. This is still due to third-harmonic interference, but with the phase shift from shot to shot coming purely from the different nonlinear phase. Turning off third-harmonic generation but including random CEP [Fig.~\ref{fig:min_direct}(c)] demonstrates this by drastically reducing the noise. Importantly, even with all effects included, the RDW pulse in time is stable [see Fig.~\ref{fig:min_direct}(d)]. This is because the peak power of the third harmonic is very low, so its influence on the time-domain pulse is weak. Note that at low pump energy, when the RDW is weak, the third harmonic can more strongly influence the shape of the filtered pulse; this is likely the cause of the small discrepancy in the timing jitter observed between the two methods compared in Fig.~\ref{fig:direct_vs_fine}.

This representative example points to a subtlety in the analysis of source noise in RDW emission: the most appropriate measure of the noise depends critically on the intended application. For instance, despite the large instability in the spectral energy density in Fig.~\ref{fig:min_direct}(a), the RDW pulse in the time domain is stable, and so a source with these noise characteristics is perfectly usable for most ultrafast experiments. If, on the other hand, the source is used directly for spectroscopy, the shot-to-shot fluctuations in the fringe phase need to be removed by averaging. An assessment purely in terms of the commonly used spectral relative intensity noise (RIN)---the standard deviation in spectral energy density normalised to its mean value---obscures this distinction \cite{adamu_noise_2020}. Furthermore, the large difference between Fig.~\ref{fig:min_direct}(a) and (b) indicates that numerical simulations of the spectral RIN which ignore the effect of random CEP are incomplete.

\subsection{Timing jitter}
\label{subsec:timing}
\begin{figure}
    \centering
    \includegraphics[width=6in]{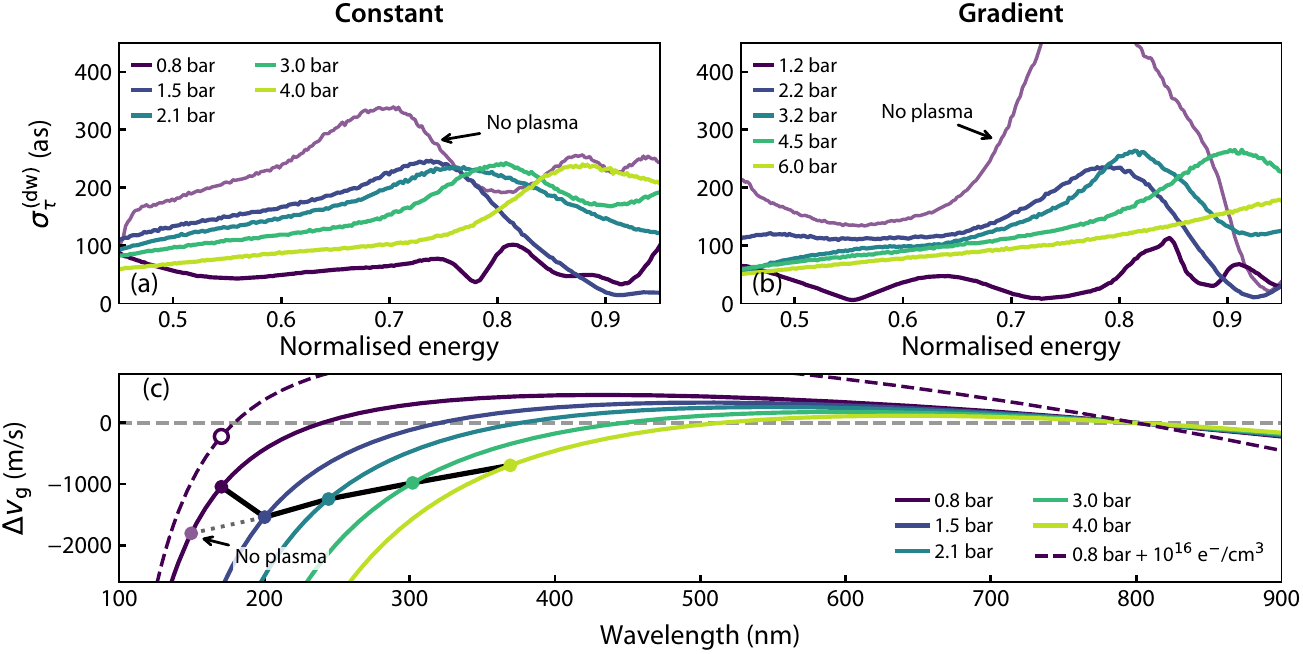}
    \caption{(a, b) Timing jitter of RDW emission as a function of mean pump energy in a \SI{1}{\m} long HCF with \SI{125}{\micro\meter} core radius filled with different pressures of helium at constant pressure (a) or with a decreasing gradient (b) and driven with \SI{7.5}{\fs} pulses at \SI{800}{\nm}. The lighter purple lines show the results for \SI{0.8}{\bar} and \SI{1.2}{\bar}, respectively, when photoionisation and plasma dynamics are ignored in the simulations. (c) Group velocity difference $\Delta v_\mathrm{g}$ with respect to \SI{800}{\nm} for the same pressures as in (a). The circles mark the value of $\Delta v_\mathrm{g}$ for the approximate wavelength of RDW emission as extracted from the simulations, with the lightened dot and dashed grey line indicating the result without plasma as for the light purple lines in (a) and (b). The dashed purple line shows $\Delta v_\mathrm{g}$ for \SI{0.8}{\bar} pressure and a free-electron density of \SI{e16}{\per\cubic\cm}.}
    \label{fig:RDWarrival_std}
\end{figure}
Figure \ref{fig:RDWarrival_std}(a) and (b) show the arrival-time jitter of the RDW pulse for the same set of parameters considered in Fig.~\ref{fig:RDWenergy_std}. The most important result here is that the jitter is generally very small compared to the pulse duration, remaining below \SI{300}{\atto\second} for all parameters---well within the requirements for ultrafast experiments with few-femtosecond resolution. In contrast to the energy noise amplification, the modelled arrival-time jitter depends on the chosen pump energy noise. We find that it is simply proportional over the entire range of energy and pressure considered here, so that halving the pump energy noise also halves the amount of timing jitter. There are two overall trends: with the exception of the lowest pressure, higher pressures (longer RDW wavelengths) generally perform better, and there is a pronounced maximum in the timing jitter as a function of pump energy.

The first of these trends is somewhat counterintuitive. As the lines in Fig.~\ref{fig:RDWarrival_std}(c) show, higher pressures lead to stronger higher-order dispersion and hence a generally larger group-velocity difference with respect to the pump pulse. However, the longer RDW wavelength resulting from these higher pressures more than compensates for this effect, as demonstrated by the circles plotted in Fig.~\ref{fig:RDWarrival_std}(c). Despite being larger in general, the group-velocity difference at the RDW wavelength is reduced. The VUV RDW generated with \SI{0.8}{\bar} pressure breaks this trend, with the timing jitter at all pump energies substantially lower than for other gas pressures. The reason for this lies in the strong influence of plasma dynamics resulting from the high pump pulse energy required at low pressure. Photoionisation by the self-compressing pulse leads to soliton self-frequency blue-shift \cite{fedotov_ionization-induced_2007,chang_influence_2011,holzer_femtosecond_2011}. Because the pump and RDW wavelengths are inversely related \cite{akhmediev_cherenkov_1995}, this shifts the RDW to longer wavelengths, where the group-velocity difference is reduced [see purple circle in Fig.~\ref{fig:RDWarrival_std}(c)]. By switching off photoionisation in the simulations, we can restore the trend observed for higher pressures, as shown by the lighter purple lines in Fig.~\ref{fig:RDWarrival_std}(a) and (b) and the lighter purple circle in Fig.~\ref{fig:RDWarrival_std}(c). This effect alone cannot account for the dramatically lower timing jitter, however, since it only reduces the group-velocity difference to the same level as for \SI{3}{\bar} pressure. It is aided by the presence of free electrons. Since the plasma dispersion is anomalous, it accelerates the VUV RDW and reduces the group-velocity difference. For an electron density of \SI{e16}{\per\cubic\cm} (an ionisation fraction of \SI{0.05}{\percent} at \SI{0.8}{\bar}), the group-velocity difference for the VUV RDW is near zero [see open purple circle in Fig.~\ref{fig:RDWarrival_std}(c)]. Free electrons are only created when the intensity is very high, so that they are localised around self-compression points. However, re-compression of the blue-shifted soliton can maintain high intensity even after the RDW is generated \cite{chang_influence_2011}. In our simulations at \SI{0.8}{\bar} pressure, the electron density remains around \SI{e16}{\per\cubic\cm} even after the first self-compression point for all energies at which an energetic RDW pulse is created.

\begin{figure}
    \centering
    \includegraphics[width=6in]{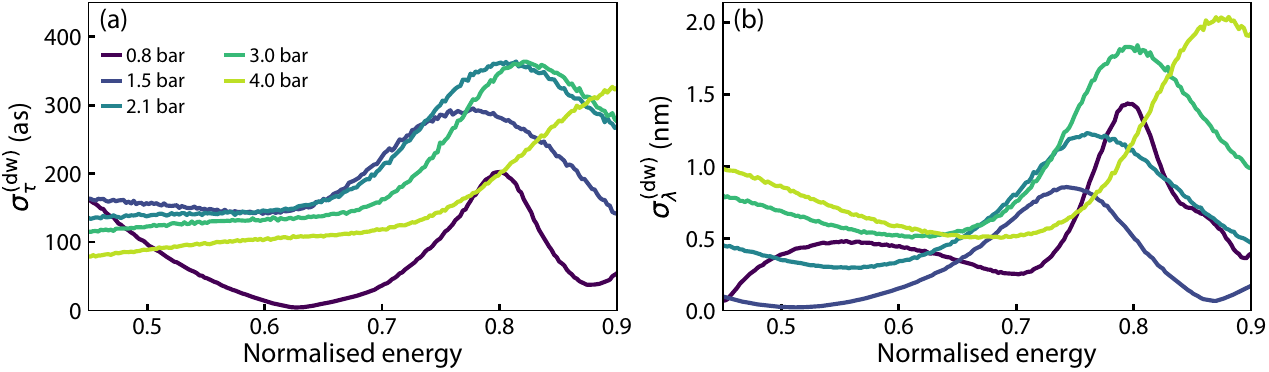}
    \caption{(a) RDW timing jitter for the same set of parameters as in Fig.~\ref{fig:RDWenergy_std}(a) and Fig.~\ref{fig:RDWarrival_std}(a) as calculated using the simple model eq.~\ref{eq:arrival_std_an}. (b) Jitter in the central wavelength of the RDW for the data in (a). }
    \label{fig:RDWarrival_an}
\end{figure}
To better understand the change in timing jitter with pump energy, we consider a simple model for the delay $\tau$ induced by the difference in group velocity of the pump pulse and the RDW pulse,
\begin{equation}
    \tau = L_\mathrm{prop}\left[\beta_1(\omega) - \beta_1(\omega_0)\right] = L_\mathrm{prop}\Delta\,,
    \label{eq:arrival_std_an}
\end{equation}
where $L_\mathrm{prop}$ is the propagation length of the RDW pulse after the generation point, $\beta_1(\omega) = \partial_\omega\beta(\omega)$ is the inverse of the group velocity, and $\omega$ and $\omega_0$ are the central frequencies of the RDW and pump pulses, respectively. The propagation length can be approximated by $L_\mathrm{prop} = L - L_\mathrm{f}$, where $L$ is the waveguide length and $L_\mathrm{f}$ is the fission length---the propagation length after which the pump pulse self-compresses and the RDW is generated. The latter is given by
\begin{equation}
    L_\mathrm{f} = \sqrt{\frac{T_0^2}{\gamma\abs{\beta_2(\omega_0)}P_0}}\,,
\label{eq:fission_length}
\end{equation}
where $T_0$ and $P_0$ are the pump pulse duration and peak power (assuming a $\sech^2$ profile), $\gamma$ is the nonlinear coefficient of the gas-filled waveguide, and $\beta_2(\omega) = \partial^2_\omega\beta(\omega)$ is the group-velocity dispersion \cite{travers_ultrafast_2011}. The pump energy enters the two factors in eq.~\ref{eq:arrival_std_an} in different ways: $L_\mathrm{prop}$ is longer for higher energy as self-compression happens more quickly ($L_\mathrm{f}$ decreases), whereas the energy-dependent wavelength of the RDW affects the group-velocity difference encoded in $\Delta$. We estimate the jitter in the arrival time by calculating the delay $\tau$ using eq.~\ref{eq:arrival_std_an} for 10000 randomly chosen pump energies around each of the mean energy values for which we show the simulated jitter in Fig.~\ref{fig:RDWarrival_std}, while finding the central wavelength of the RDW in the same manner as for our choice of window function.

Figure \ref{fig:RDWarrival_an}(a) shows the predictions for the timing jitter from this simple model. The model closely reproduces the magnitude of the timing jitter, which indicates that the group-velocity difference in combination with a changing fission length is indeed the dominant factor. The general trend of longer-wavelength RDWs at higher pressures being more stable is also captured. Importantly, while the jitter is generally small for the lowest pressure, the simple model does \emph{not} reproduce the extent to which the jitter is suppressed, especially at higher energy. This is because only the refractive index of the neutral gas is considered in eq.~\ref{eq:arrival_std_an}; the effect of the plasma is not included. Figure \ref{fig:RDWarrival_an}(b) shows the jitter in the central wavelength of the RDW. This is extracted from the simulations in an analogous manner to the energy and timing, using the first moment of the filtered spectrum. The model results reveal that the central-wavelength jitter is the origin of the pronounced maximum observed in Fig.~\ref{fig:RDWarrival_std}(a) and (b). For certain energy ranges, the RDW very rapidly blue-shifts with increasing energy, and the group-velocity difference increases correspondingly. Note that the maxima in central-wavelength and timing jitter appear close to the minima in overall energy noise [Fig.~\ref{fig:RDWenergy_std}(a) and (b)]. Thus it may be preferable to tune the generation parameters slightly away from this region for the best combination of stable pulse energy and timing.

\subsection{Effect of the pump pulse duration}
\label{subsec:pumpduration}
\begin{figure}
    \centering
    \includegraphics[width=6in]{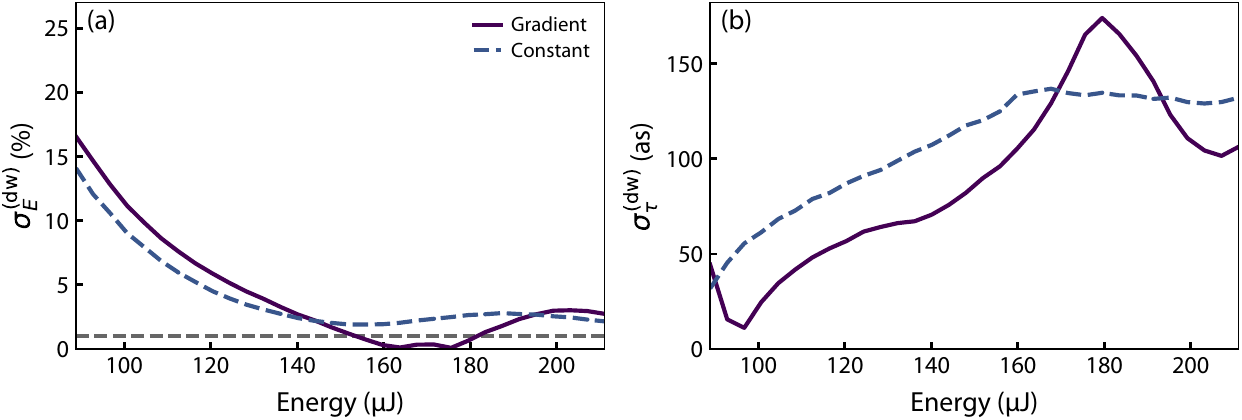}
    \caption{Energy (a) and timing (b) noise for the same conditions as shown in Figs.~\ref{fig:RDWenergy_std} and \ref{fig:RDWarrival_std} but assuming \SI{1}{\percent} fluctuations in the pump pulse duration. The solid lines show results for a gradient from \SI{3.2}{\bar} to vacuum and the dashed lines for a constant helium pressure of \SI{2.1}{\bar}.}
    \label{fig:duration_energy}
\end{figure}
As exemplified by the expression for the fission length in eq.~\ref{eq:fission_length}, the dynamics of soliton self-compression depend on the pump pulse duration as well as the energy. The extent to which the duration of femtosecond laser pulses fluctuates varies greatly with the generation method, as does the degree of correlation with the pump pulse energy. For instance, in nonlinear pulse compression, the output bandwidth is coupled to the input energy, but whether the output duration increases or decreases with larger bandwidth depends on the details of the subsequent dispersion compensation. Similarly, the duration of laser pulses can change due to fluctuations in bandwidth, the spectral phase, or both, again depending on the details of the generation mechanism. We therefore treat the pulse duration fluctuations independently and only consider a change in the duration of a perfectly compressed input pulse (that is, changes in bandwidth but not spectral phase).

Figure~\ref{fig:duration_energy} shows the energy and timing noise induced by a \SI{1}{\percent} fluctuation in the pump pulse duration for one of the sets of parameters shown before. The behaviour with changing pump energy is strikingly similar to that observed when the energy fluctuates instead, particularly for the RDW energy noise. At low energies, the RDW energy fluctuates severely, but as the pump energy is increased and RDW generation saturates, the energy noise improves greatly; for the case of a pressure gradient, it falls close to zero. 

The induced timing jitter is of a similar magnitude to that shown in Fig.~\ref{fig:RDWarrival_std}, despite the smaller relative pump noise in our model (\SI{1}{\percent} in duration as compared to \SI{2}{\percent} in energy). This is because the fission length depends more sensitively on the duration than on the energy---taking into account that the peak power $P_0$ is proportional to $E_p/T_0$, where $E_p$ is the pump energy, the fission length scales as $T_0^{3/2}$ and $E_p^{-1/2}$, respectively, when the waveguide parameters $\gamma$ and $\beta_2$ are fixed. Here, however, the different possible relationships between pump duration and energy become important: for a pump source in which $E_p$ and $T_0$ are anti-correlated (that is, in which higher pump energy corresponds to shorter pump pulses), these two coupled noise sources will result in larger energy and timing noise in the dispersive wave. In the opposite case, the effect of an increase in pump energy is partly counteracted by a corresponding increase in pump pulse duration, resulting in lower overall noise.

\subsection{Overall energy scaling}
\begin{figure}
    \centering
    \includegraphics[width=6in]{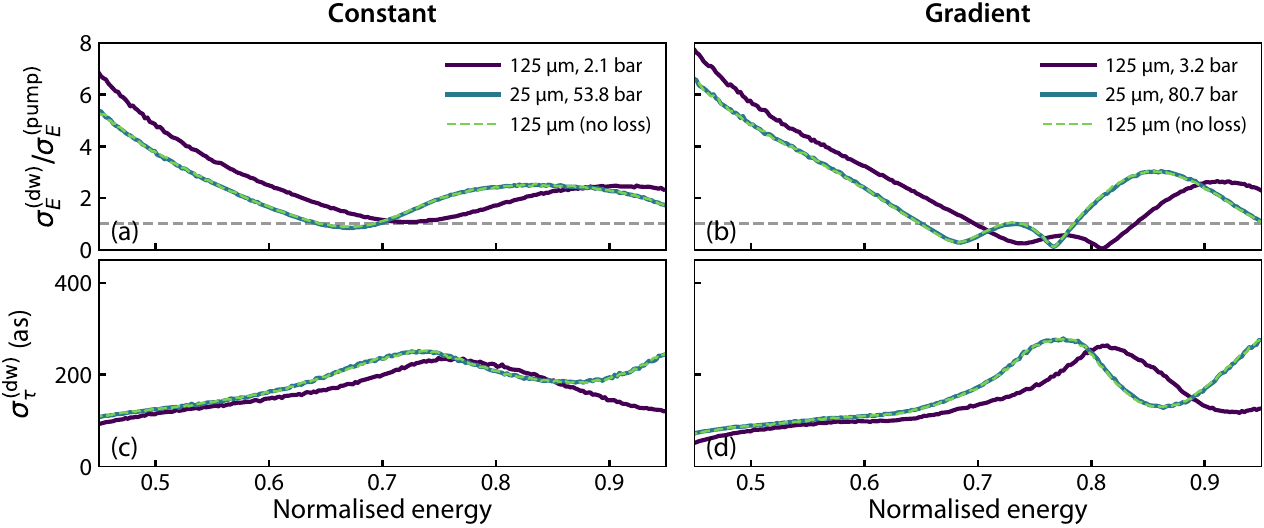}
    \caption{Effect of overall energy scaling on the (a, b) energy and (c, d) arrival-time noise of RDW emission for both a constant pressure (left column) and a decreasing gradient (right column).}
    \label{fig:energy_scaling}
\end{figure}

Nonlinear optical effects in gases, including soliton self-compression and resonant dispersive wave emission, can be arbitrarily scaled in energy by appropriately scaling the gas density and the longitudinal as well as transverse dimensions (here, waveguide length and core radius, respectively) \cite{heyl_scale-invariant_2016,travers_high-energy_2019,brahms_high-energy_2019}. In this way, RDW emission with the same temporal and spectral properties can in principle be obtained in both small-core microstructured fibres \cite{joly_bright_2011,travers_ultrafast_2011,mak_tunable_2013,ermolov_supercontinuum_2015,kottig_generation_2017} and large-core hollow capillary fibres \cite{travers_high-energy_2019,brahms_high-energy_2019,brahms_infrared_2020}. In Fig.~\ref{fig:energy_scaling} we compare the noise characteristics of these two regimes. The data for the HCF with \SI{125}{\micro\meter} core radius is the same as shown in Figs.~\ref{fig:RDWenergy_std} and \ref{fig:RDWarrival_std}. We simulate propagation in a hollow-core PCF with a core radius of \SI{25}{\micro\meter} using the same model but assuming negligible waveguide loss. The core radius scaling by a factor of 5 results in 25 times lower overall energy, shorter length and higher density. (Note that the \emph{pressure} is not scaled by exactly 25 when taking into account the full equation of state for helium \cite{bell_pure_2014}.)

The overall trend of the energy and arrival-time noise is very similar for both the small and large core, but the curves do not agree exactly. This is due to the difference in propagation loss---the transmission of the \SI{1}{\meter} HCF is $\sim\!\SI{87}{\percent}$ at \SI{800}{\nm}---which the general scaling law does not account for. Switching off the waveguide loss in the large-core simulations confirms this hypothesis. As the green dashed line in Fig.~\ref{fig:energy_scaling} shows, without loss, the noise characteristics are exactly the same despite the large difference in overall energy and the physical size of the system.

\section{Conclusions}
In summary, we have investigated the energy and arrival-time noise of RDW emission in gas-filled hollow waveguides for a range of generation conditions and with a focus on applications in ultrafast science. There are two key results from this study. The first is that, when saturated, RDW emission can be as stable as or even more stable than the pump source driving the process, while still generating ultrashort pulses. This is the case for a wide range of RDW wavelengths, so that the full flexibility of RDW-based sources can be exploited. The second key result is that even for a conservative estimate of the pump energy noise, the timing jitter induced by the wavelength-dependent group velocity and resulting pulse walk-off remains well below one femtosecond. As many ultrafast laser systems, even those including nonlinear compression, perform far better than the \SI{2}{\percent} energy noise we have considered here, a timing jitter below \SI{100}{\as} should be achievable in experiments. There are clear wavelength-dependent trends in the timing jitter and strong plasma generation can dramatically improve the stability by reducing the group-velocity mismatch. Instabilities in the pump pulse duration have a similar effect to those in pump energy, though with larger timing jitter for the same relative magnitude of noise. Finally, by scaling the model to parameters similar to experiments in anti-resonant hollow-core fibres, we have demonstrated that our results are widely applicable in those systems as well. We expect that our results will aid in the design and implementation of versatile, stable ultrafast light sources based on RDW emission.

\section*{Acknowledgments}
This work was funded by the European Research Council under the European Union’s Horizon 2020 Research and Innovation program: Starting Grant agreement HISOL, No.~679649; Proof of Concept Grant agreement ULIGHT, No.~899900.

\bibliography{bibliography}

\end{document}